\documentclass[%
aip,%
amsmath,amssymb,floatfix,
reprint,%
twocolumn,%
longbibliography,
]{revtex4-1}

\usepackage{graphicx}
\usepackage{dcolumn}
\usepackage{bm} 
\usepackage[usenames,dvipsnames]{xcolor}
\usepackage{soul}

\usepackage[left= 1.53cm, right = 1.53cm, top= 2.cm, bottom = 2.cm]{geometry}  

\usepackage[english]{babel} 

\usepackage{amsmath,amsthm,latexsym,amssymb,amsfonts,epsfig}
\usepackage{mathtools}
\usepackage{mathrsfs}

\usepackage{xcolor}
\definecolor{OliveGreen}{rgb}{0,0.6,0}

\newcommand{\vect}[1]{\boldsymbol{#1}}

\newcommand{\R}{\vect{r}}

\newcommand{\Intd}{\mathrm{d}}

\def\XXint#1#2#3{{\setbox0=\hbox{$#1{#2#3}{\int}$}
     \vcenter{\hbox{$#2#3$}}\kern-.5\wd0}}

\newcommand{\rhatij}{\vect{\hat{r}}_{ij}}

\newcommand{\eX}{\vect{\hat{e}}_x}
\newcommand{\eY}{\vect{\hat{e}}_y}
\newcommand{\eZ}{\vect{\hat{e}}_z}

\newcommand{\bT}{\mu}

\newcommand{\rC}{r_\mathrm{C}}

\newcommand{\sqt}{\sqrt{2}}
\newcommand{\tS}{t_\mathrm{S}}

\newcommand{\Ref}[1]{{\textcolor{black}{#1}}}

\usepackage[colorlinks=true,citecolor=OliveGreen,linkcolor=blue,urlcolor=blue]{hyperref}

\usepackage{epstopdf}

\usepackage{csquotes}
\usepackage{setspace}

\usepackage{multirow}

\usepackage{type1ec} %

\usepackage{gensymb} 

\allowdisplaybreaks

\usepackage{makecell}

\begin{document}


\title{Theory of active particle penetration through a planar elastic membrane}

\author{Abdallah Daddi-Moussa-Ider} 
\email{abdallah.daddi.moussa.ider@uni-duesseldorf.de}
\affiliation
{Institut f\"{u}r Theoretische Physik II: Weiche Materie, Heinrich-Heine-Universit\"{a}t D\"{u}sseldorf, Universit\"{a}tsstra\ss e 1, 40225 D\"{u}sseldorf, Germany}

\author{Benno Liebchen}
\affiliation
{Institut f\"{u}r Theoretische Physik II: Weiche Materie, Heinrich-Heine-Universit\"{a}t D\"{u}sseldorf, Universit\"{a}tsstra\ss e 1, 40225 D\"{u}sseldorf, Germany}

\affiliation
{Theorie Weicher Materie, Fachbereich Physik, Technische Universit\"{a}t Darmstadt, Hochschulstra\ss e 12, 64289 Darmstadt, Germany}

\author{Andreas M. Menzel}
\affiliation
{Institut f\"{u}r Theoretische Physik II: Weiche Materie, Heinrich-Heine-Universit\"{a}t D\"{u}sseldorf, Universit\"{a}tsstra\ss e 1, 40225 D\"{u}sseldorf, Germany}

\author{Hartmut L\"{o}wen}
\email{hartmut.loewen@uni-duesseldorf.de}
\affiliation
{Institut f\"{u}r Theoretische Physik II: Weiche Materie, Heinrich-Heine-Universit\"{a}t D\"{u}sseldorf, Universit\"{a}tsstra\ss e 1, 40225 D\"{u}sseldorf, Germany}

\date{\today}

\begin{abstract}

With the rapid advent of biomedical and biotechnological innovations, a deep understanding of the nature of interaction between nanomaterials and cell membranes, tissues, and  organs, has become increasingly important.
Active penetration of nanoparticles through cell membranes is a fascinating phenomenon that may have important implications in various biomedical and clinical applications.
Using a fully analytical theory supplemented by particle-based computer simulations, the penetration process of an active particle through a planar two-dimensional elastic membrane is studied.
The membrane is modeled as a self-assembled sheet of particles, uniformly arranged on a square lattice.
A coarse-grained model is introduced to describe the mutual interactions between the membrane particles.
The active penetrating particle is assumed to interact sterically with the membrane particles.
State diagrams are presented to fully characterize the system behavior as functions of the relevant control parameters governing the transition between different dynamical states.
Three distinct scenarios are identified.
These compromise trapping of the active particle, penetration through the membrane with subsequent self-healing, in addition to penetration with permanent disruption of the membrane.
\Ref{The latter scenario may be accompanied by a partial fragmentation of the membrane into bunches of isolated or clustered particles} and creation of a hole of a size exceeding the interaction range of the membrane components.
It is further demonstrated that the capability of penetration is strongly influenced by the size of the approaching particle relative to that of the membrane particles.
Accordingly, active particles with larger size are more likely to remain trapped at the membrane for the same propulsion speed.
Such behavior is in line with experimental observations.
Our analytical theory is based on a combination of a perturbative expansion technique and a discrete-to-continuum formulation.
It well describes the system behavior in the small-deformation regime.
Particularly, the theory allows to determine \Ref{the membrane displacement of the particles in the trapping state.}
Our approach might be helpful for the prediction of the transition threshold between the trapping and penetration in real-space experiments involving motile swimming bacteria or artificial active particles.

\end{abstract}

\maketitle

\section{Introduction}

As one of the most fundamental components in biological systems, the cell membrane 
defines and protects the cell and is selectively permeable for ions and organic molecules, allowing to control the movement of required chemicals into the cell and of unwanted products out of the cell. 
It is now possible not only to reassemble cell membranes 
artificially \cite{Budin2011}, but also to design synthetic membranes with properties tailored
to the needs of 21st centuries societies \cite{Osada1992,Belfort2012, brown17}. 
In fact, synthetic membranes are now routinely used already for applications from water purification \cite{Fane2015,Kocsis2018} to dialysis \cite{Vienken1999,Klinkmann1995} and can be regarded as a paradigmatic success of biomimetics~\cite{davis07, kim12synthetic, jun10}. 
Future perspectives for the usage of synthetic membranes involve problems like 
targeted gene and drug delivery to (cancer) cells \cite{langer88, verma08,Yang2010,lin10,verma10,Nel2009,wang2012cellular,shang14, monzel17, lisse17,muller18,xuan18self, wang19gold} or, more generally, the delivery of cargo to 
the interior of synthetic droplets, requiring a precise understanding of the interaction of motile particles 
with synthetic and biological membranes. 
Evidence from previous studies has shown that the physical uptake by living cells is strongly affected by the particle and membrane physicochemical and functional properties~\cite{chithrani06, yang10, dos11, parodi13, daddi16, juenger15, grafe16}.

While membranes comprising active inclusions, 
e.g.\ in the form of embedded proteins creating a stress on the membrane, have been studied for decades
\cite{Ramaswamy2001,yoon11,Lacoste2014}, the penetration of active particles 
through the membrane is less explored \cite{Yang2010,Mitragotri2009} with the few existing studies focusing on nano- and biotechnology perspectives. 
In particular, penetration of nanoparticles through a membrane has been studied using dissipative particle dynamics simulations, focusing on effects of particle shape 
\cite{Yang2010} and surface-structure \cite{Li2012}.
In addition, molecular dynamics simulations have been employed to investigate the penetration of fullerens through lipid membranes~\cite{Wong2008}.
Recent studies have also explored interactions of active particles with membranes, from a more physical point of view, but did not focus on particle penetration \cite{Junot2017,Marconi2017,Costanzo2014}.
For a 1D membrane, we have recently performed a corresponding investigation~\cite{daddi19penetration}.

Conversely to most of the above works, here we explore the penetration of an active particle through a 2D synthetic membrane from a physics perspective, aiming at predicting overall properties such as the membrane shape or the parameter domain leading to penetration starting
from coarse microscopic details.
\Ref{We focus on a minimal model membrane that can be realized in principle using as building blocks microparticles interacting via elastic forces~\cite{boal92, lidmar03, noguchi05, fedosov10, fortsch17, laumann17, laumann19}.
Other types of interactions, such as, dipolar interactions may be considered as well to model self-assembled chains and sheets~\cite{froltsov03, Barry2010, Ewerlin2013, Messina2014, kaiser15, Guzman-Lastra2016, yener16, Spiteri2017, Messina2017, peroukidis16, deissenbeck18, Garcia-Torres2018, oguz18}.}
To predict the state diagram, informing us about the parameter domains where particles can penetrate through the membrane and where they cannot, we systematically derive a continuum description of the membrane. 
We compare our results with particle-based computer simulations, finding close quantitative agreement regarding the transition between trapping and penetrating states, membrane shape and dynamics. 
Our analytical closed-form expressions might help to predict the properties of synthetic membranes, e.g.\ regarding 
the speed and size of particles which will be able to penetrate through them.

Below, we first define our model (Sec.~\ref{sec:systemSetup}), followed by a
brief discussion of the relevant parameters and of the
2D membrane dynamics as induced by the active particle approaching it
(Sec.~\ref{sec:stateDiagram}).
Here, besides a trapping state in which the membrane is deformed in the
final state and does not allow the particle to pass, we find
two scenarios of penetration.
The first of these corresponds to the
particle breaking through the membrane,
followed by a complete self-healing of the membrane,
which might be the desired behavior when delivering cargo towards a
synthetic droplet or a healthy cell. 
The second scenario of penetration occurs mainly for larger particles, creating a hole in the membrane with a size exceeding the interaction range of the membrane components.
This situation is accompanied by a partial fragmentation of the membrane structure into isolated particles.
Following this qualitative discussion, we systematically explore the corresponding
state diagram using numerical simulations, showing for which parameter
combinations which of these three states prevails, and we develop a detailed
analytical theory (Sec.~\ref{sec:analytischeTheorie}).
The latter is able to predict essentially the
entire state diagram as well as the shape and the dynamics of the
membrane, in close quantitative agreement with our simulations.
Interestingly, the transition between self-healing and non-healing
states is sharp, suggesting that there is a critical size for particles
that pass a membrane by causing significant damage. 
This also suggests that if one were to
permanently damage the membrane in our minimal model (and perhaps
similarly in practice to treat cancer cells), one needs to use particles with a certain
minimal size.
Finally, concluding remarks summarizing our findings are contained in Sec.~\ref{sec:conclusions}.

\section{System setup}
\label{sec:systemSetup}

\begin{figure}
	\centering
	\vspace{-1cm}
		\includegraphics[scale=0.45]{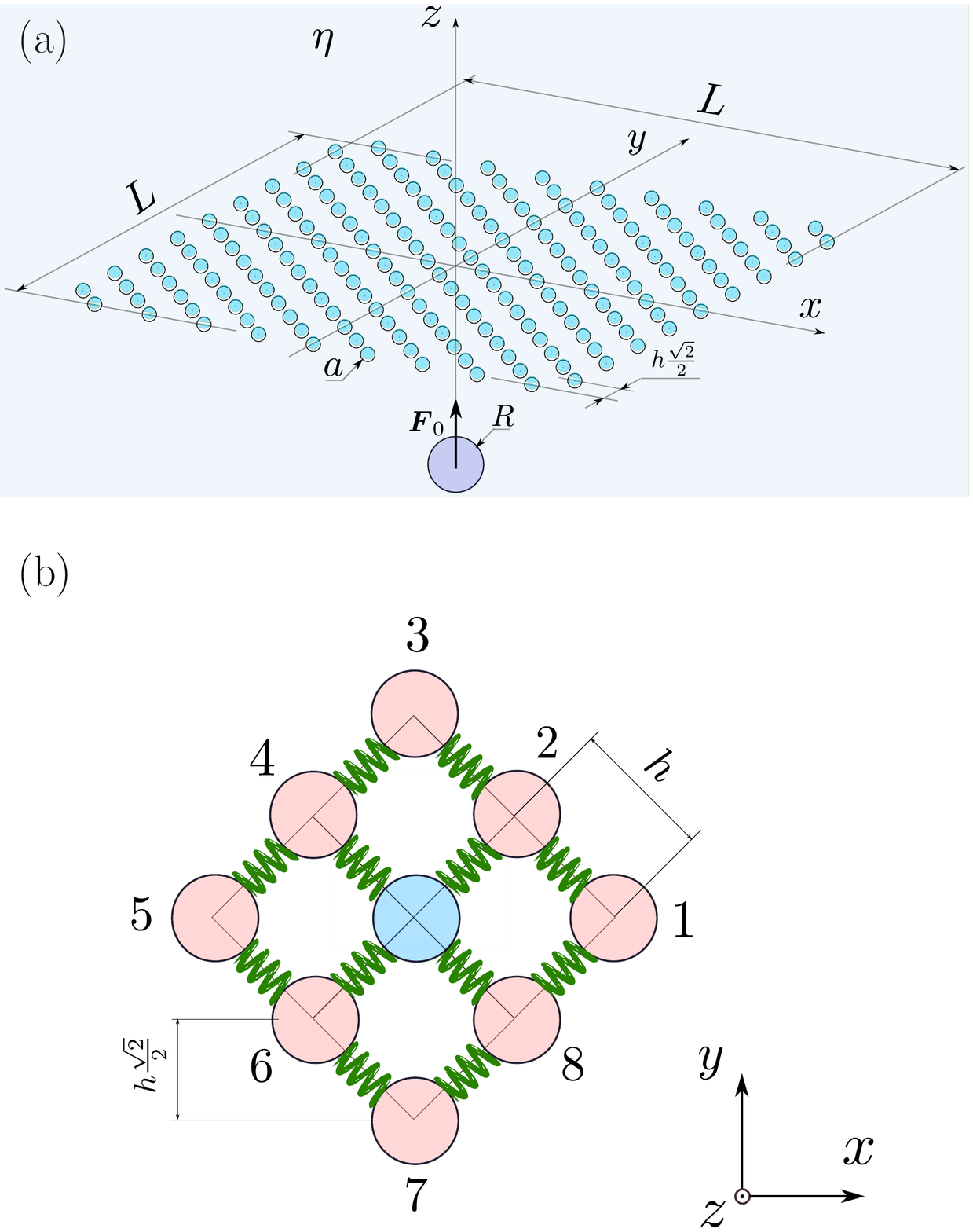}
		\vspace{-1cm}
		\caption{(Color online) Graphical illustration of the system setup.
		$(a)$ An active particle of radius~$R$ moving through an effective driving force~$\vect{F}_0$ toward a two-dimensional membrane composed of~$N$ particles of radius~$a$.
		The membrane particles are initially arranged on a square lattice of dimension~$L\times L$ and spacing~$h$, rotated by~$45\degree$ around the~$z$~axis,
		the latter oriented normal to the plane of the membrane.
		The membrane is centered about the origin and clamped at its periphery.
		Periodic boundary conditions are imposed in both~$x$ and~$y$ directions.	
		We assume that the membrane particles are subject to steric and elastic pairwise interactions.
		The system is fully immersed in a Newtonian viscous fluid of shear viscosity~$\eta$.
		$(b)$ Schematic illustration of the lattice structure composing the model membrane.
		For future reference, the eight nearest neighbors of the particle at the center of the lattice are identified by numbers (1--8).
		Elastic springs are also inserted along the lattice diagonals but are not displayed here for reasons of clarity.
		}
		\label{illustration}
\end{figure}

We examine the penetration mechanism of a non-fluctuating membrane by an active particle moving under the action of a constant propulsion force~$\vect{F}_0$.
\Ref{We assume that the persistence length of the self-propelling active particle is larger than the distance initially separating the particle from the membrane.
Correspondingly, we focus on the limiting case of vanishing rotational diffusion.
This implies that the particle essentially moves along a straight trajectory without changing its orientation.~\cite{tenHagen11, wittkowski12, kaiser12, wensink12, kummel13, tenhagen14, tenHagen15, speck16, degraaf16, liebchen17b, hoell17, daddi18nematic}.}
The active particle may represent a swimming microorganism~\cite{lauga09, zottl16, lauga2016ARFM,elgeti15, bechinger16, illien17} or an artificial microrobot that can be manipulated by controlled external fields~\cite{gao13, gao14, scholz18, scholz18delay}.

In our model, the membrane is composed of~$N$ identical \Ref{spherical beads (or vertices)} of radius~$a$, uniformly arranged on a square lattice of size~$L\times L$, rotated by~$45\degree$ around the axis~$z$, the latter directed normal to the membrane, as schematically illustrated in Fig.~\ref{illustration}.
We denote by~$h$ the lattice spacing after initialization.
The membrane is immersed in a Newtonian fluid, characterized by a constant dynamic viscosity~$\eta$.
We support the membrane at its periphery (the particle displacements are zero for~$x,y = \pm L/2$) and assume periodic boundary conditions in the transverse directions~$(x,y)$.
Moreover, we suppose that the mutual interactions between the membrane particles are pairwise additive and described by forces that depend only on the difference of coordinates of each two neighboring particles.
\Ref{Representing the membrane as a collection of spherical beads arranged on vertices has extensively been employed as a coarse-grained model for cell membranes, see, for instance Refs.~\onlinecite{boal92, lidmar03, noguchi05, fedosov10, fortsch17, laumann17, laumann19}.}

Typically, various types of interactions may occur among membrane particles including steric and elastic interactions.
For instance, steric interactions can be imposed by membrane phospholipids chains and other biomolecules~\cite{adey77, jiang07, demirors15}, whereas intermolecular coupling between the lipid bilayer and the cytoskeleton network gives rise to elastic interactions~\cite{gov07, pivkin08, peng13, mejean13}.
Accordingly, the total potential energy of the membrane here is written as a sum of \Ref{two distinct contributions} as
\begin{align}
			\mathcal{U} &= 4\epsilon \sum_{\substack{i,j=1 \\ j< i}}^{N} N_{ij}  
				 \left(
				 	\left(\frac{\sigma}{r_{ij}}\right)^{6}
				 	\left( \left(\frac{\sigma}{r_{ij}}\right)^{6}
				 	- 1 \right) + \frac{1}{4} \right) \notag \\
				 &\quad+ \frac{k}{2} \sum_{i=1}^{N} \sum_{\substack{j \in \mathcal{N}(i) \\ j< i}}
				  \left( r_{ij} - \xi {r_0}_{ij} \right)^2 \, , \label{potentialEnergy}
	\end{align}
wherein $r_{ij} = |\R_{ij}|$ is the distance between particles~$i$ and~$j$, $\R_{ij} = \R_i-\R_j$, and $\rhatij = \R_{ij} /r_{ij}$ is the corresponding unit distance vector.
In addition, $\epsilon$ is an energy scale associated with the Weeks-Chandler-Anderson (WCA) pair-potential~\cite{weeks71}, $\sigma = 2a$ is the diameter of the particles, $N_{ij} = H \left( r_\mathrm{C} - r_{ij} \right)$, with~$H(\cdot)$ denoting the Heaviside step function, and $r_\mathrm{C} = 2^{1/6} \sigma$ is a finite cutoff distance beyond which the steric interactions energy vanishes.
Furthermore, $k$~is the elastic constant of the harmonic springs coupling each particle to its four nearest and four next-nearest neighbors, ${r_0}_{ij}$ is the rest length of the springs, and~$\xi \in (0,1]$ is a prestress parameter. 
Here, we use the notation~$\mathcal{N}(i)$ to denote the set of nearest and next-nearest neighbors of the~$i$th membrane particle.
\Ref{
For real cell membranes, the lattice spacing~$h$ may, for instance, be viewed as an average distance between cytoskeleton-bilayer connection sites.
In addition, the elastic constant $k$ may be connected to the shear modulus of the cytoskeleton network, the order of magnitude of which is about~$10^{-6}$~N/m.}

For the sake of simplicity, we neglect all possible hydrodynamic interactions between particles.
Moreover, we assume that the particles are small enough or sufficiently matched in density to the surrounding fluid for the influence of gravity to be neglected, and large enough for the effect of thermal fluctuations to be neglected.
\Ref{In addition, we assume throughout this work that the size of the active particle is comparable or larger than that of the membrane particles.}

The corresponding interaction force acting on the~$i$th membrane particle is obtained by differentiating the potential energy described by Eq.~\eqref{potentialEnergy} with respect to the particle position~\cite{babel16} as $\vect{F}_i = -\partial \mathcal{U} / \partial \vect{r}_i$.
Accordingly,
\begin{align}
			\vect{F}_i  &= 48\epsilon \sum_{\substack{j=1 \\ j\neq i}}^{N} \frac{N_{ij}}{r_{ij}}
					  		\left(\frac{\sigma}{r_{ij}}\right)^6
							 \left( \left(\frac{\sigma}{r_{ij}}\right)^6- \frac{1}{2} \right)
							 \rhatij \notag \\
				    		&\quad+ k \sum_{j\in \mathcal{N}(i)}
				    	 \left( \xi {r_0}_{ij} - r_{ij} \right) \rhatij \, . \label{force}
\end{align}

At small length scales, aqueous systems are characterized by small Reynolds numbers, so that viscous forces dominate over inertial forces.
The resulting overdamped dynamics can therefore be adequately described within the framework of linear hydrodynamics~\cite{happel12, kim13}.
Accordingly, the translational velocity of the membrane particles~$\vect{V}_i$ is linearly coupled to the forces acting on their surfaces via the hydrodynamic mobility functions~\cite{swan07, swan10, balboa17, driscoll18}.
The latter are second-order tensors, which simply reduce to scalar quantities when considering motion in an unbounded medium and neglecting the fluid-mediated hydrodynamic interactions between the particles.
Specifically,
\begin{equation}
		\vect{V}_i = \frac{\Intd \R_i}{\Intd t} = \bT \left( \vect{F}_i + \vect{F}^\mathrm{ext}_i \right) \, ,  \label{TranslationalDef}
	\end{equation}
where $\mu$ denotes the translational self-mobility functions of the membrane particles.
This is given by the usual Stokes formula for an isolated sphere in an infinite fluid domain as $\mu = 1/(6\pi\eta a)$.
In addition, $\vect{F}_i^\mathrm{ext}$ represent the external force exerted by the active particle due to the steric interactions with the membrane particles. 
These pair interactions are modeled via a soft repulsive WCA potential as in Eq.~\eqref{potentialEnergy} for which~$\sigma = R+a$, with~$R$ denoting the radius of the active particle.

We introduce at this point an additional cutoff length~$\ell$ beyond which the elastic interactions are set to zero.
Accordingly, the elastic potentials are also shifted to this cutoff length, so as to ensure that the resulting forces are continuous.

\section{Trapping, penetration, and self-healing}
\label{sec:stateDiagram}

\begin{center}
	\begin{table}
		\def\arraystretch{2.5} 
		\Ref{
		\begin{tabular}{|c|c|c|}
					\hline
					\makecell{~Dimensionless  Number~} & ~~Expression~~ & Denomination \\
					\hline
					\hline
					~$E$~ & ~$\frac{a F_0}{\epsilon}$~ & Reduced activity \\
					\hline
					~$\kappa$~ & ~$ \frac{akh}{2\epsilon}$~ & Reduced stiffness \\
					\hline
					~$\delta$~ & ~$\frac{R}{a}$~ & ~Size ratio~ \\
					\hline
					~$\lambda$~ & ~$\frac{\ell}{h}$~ & ~\makecell{Scaled \\ cutoff distance}~\\
					\hline
					~$P_0$~ & ~$ \frac{E}{\kappa} = \frac{2F_0}{kh} $ & ~Admittance~ \\
					\hline\hline
		\end{tabular}
		}
		\caption{\Ref{Dimensionless numbers characterizing the system in the trapping and penetrating states.}}
		\label{dimensionlessNumbers}
	\end{table}
\end{center}

Having introduced a model for our membrane and derived the corresponding equations governing the translational dynamics of the particles composing the membrane, we next study in detail the dynamical states emerging from the interaction between an active particle propelling toward the membrane.
For that purpose, we solve numerically the set of ordinary differential equations in time given by Eqs.~\eqref{force} and \eqref{TranslationalDef} using a standard Euler scheme with adaptive time stepping~\cite{press92}. 
Before the active particle starts to interact with the membrane particles, we assume that the lattice spacing~$h$ is identical to the cutoff length scale~$\rC$ associated with the WCA pair potential.
In addition, we assume that the rest length of the elastic springs is equal to the initial interparticle separation, i.e., ${r_0}_{ij}=h$ for the pairs of particles located along the lattice axes, and ${r_0}_{ij} = \sqt h$ for the pairs along the diagonal.
Under these conditions, the membrane is initially at equilibrium, on account of the periodic boundary conditions imposed along the transverse directions~$(x,y)$. 
We further mention that requiring~$h=\rC$ is equivalent to considering a constant ratio $h/a=2^{7/6}$.
Unless stated otherwise, we consider throughout the present article a membrane composed of~$N=450$ particles and set the prestress parameter as~$\xi=0.9$.

\Ref{
We now introduce the reduced activity
\begin{equation}
	E = \frac{a F_0}{\epsilon} \, , 
\end{equation}
which represents a balance between the magnitude of the active driving force~$F_0$ and the steric forces at particle contact.
Further, we define the reduced stiffness
\begin{equation}
	\kappa = \frac{akh}{2\epsilon} \, , 
\end{equation}
which quantifies the importance of the elastic forces relative to the steric forces.
The prefactor one half follows from theoretical considerations as will be shown in the sequel.
In addition, we introduce the size ratio
\begin{equation}
	\delta = \frac{R}{a}
\end{equation}
to denote the radius of the active particle relative to that of the membrane particles.
Finally, we define 
\begin{equation}
	\lambda = \frac{\ell}{h} \, , 
\end{equation}
as a scaled cutoff distance beyond which the elastic interactions vanish.
Unless otherwise stated, we set~$\lambda = 3/2 > \sqt$, such that the pair-interactions between the membrane particles are restricted to the four nearest and four next-nearest neighbors only. 
} \Ref{Choosing larger values of $\lambda$ hardly changes our results.}

\begin{figure*}
\begin{center}
\includegraphics[scale=1]{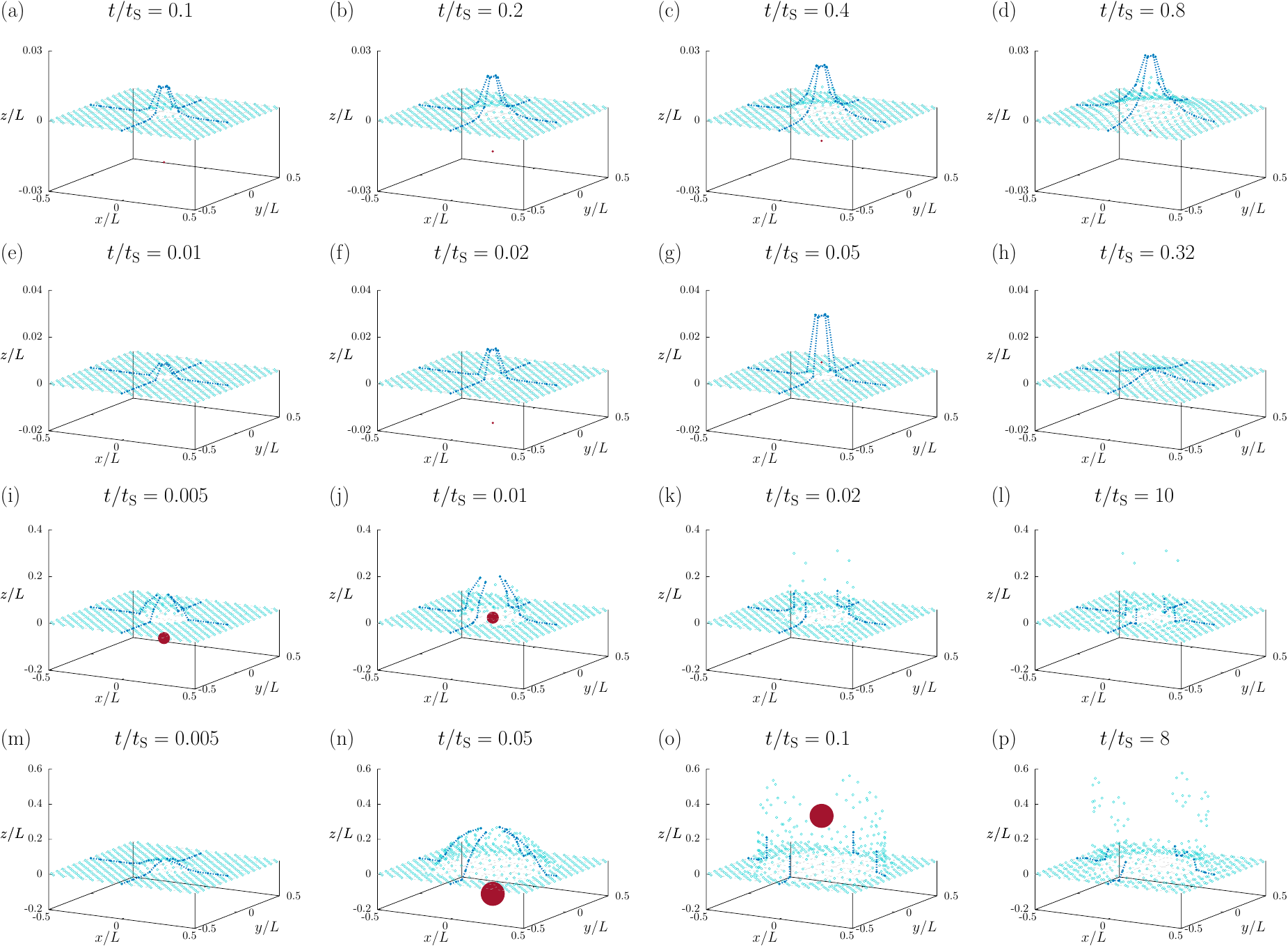}
\end{center}
\caption{(Color online) 
Snapshots of particle-based computer simulations illustrating the membrane conformation in the trapping and penetration states at different time intervals scaled by the unit simulation time~$\tS = \eta L^3/\epsilon$.
At time~$t=0$, the active particle begins interacting with the membrane particles (cyan circles). 
Here, \Ref{we set the reduced stiffness} \Ref{to} \Ref{$\kappa=10^{-2}$} in all these simulations.
For clarity, the membrane particles located in the planes~$x=0$ and~$y=0$ within the interaction range of the elastic potentials are shown as blue disks linked by dashed lines.
The first row [panels~$(a)-(d)$] displays the membrane dynamics in the trapping state, for a size ratio~$\delta=1$ and a reduced activity~\Ref{$E=10^{-2}$.}
Since the driving force is not strong enough compared to the membrane restoring forces, the active particle remains trapped near the membrane.
The second row [panels~$(e)-(h)$] represents the time frames during penetration with subsequent self-healing, for $\delta=1$ and \Ref{$E=10^{-1.5}$.}
In this state, the membrane recovers its initial planar shape after the active particle has passed through it. 
Next, the third row [panels~$(i)-(l)$] contains the frame series during penetration without self-healing, \Ref{for~$\delta=7$ and $E=\sqrt{10}$.}
The membrane remains permanently damaged after penetration as the mutual distance between the four depicted fragmented particles becomes larger than \Ref{the scaled cutoff distance~$\lambda$.}
The bottom row [panels~$(m)-(p)$] further illustrates the penetration state without self-healing, for~\Ref{$\delta = 13$ and $E=\sqrt{10}$.}
Due to the relatively large size of the active particle, 
\Ref{the membrane is partially fragmented around its center into four clusters of particle triplets and four clusters of particle sextuplets,} 
the distance between \Ref{these particles} and the remaining of the membrane is larger than the cutoff length~$\ell$.
The red disks represent the positions of the active particle, which are out of the field of view in some panels.
Because of the pronounced difference between the scales along the lateral and normal directions, the particles and their shapes are not plotted to scale.
}
\label{IllusStats}
\end{figure*}

\begin{figure}
\begin{center}
\includegraphics[scale=1]{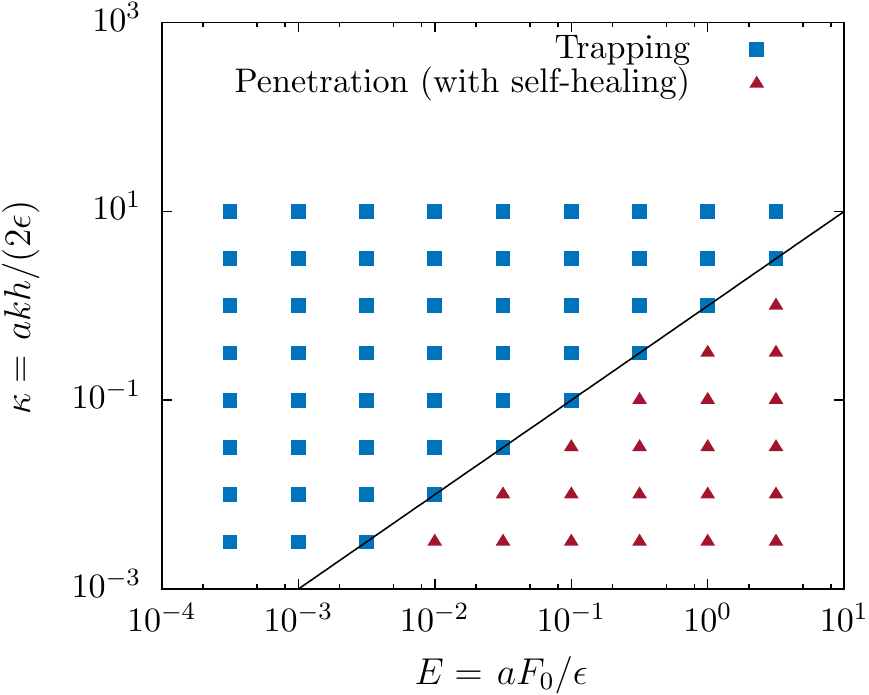}
\end{center}
\caption{\Ref{(Color online)
State diagram of membrane penetration and trapping of an active particle in the parameter space~$(\kappa, E)$.
Symbols correspond to the dynamical state resulting from numerically integrating the governing equations of motion stated by Eqs.~\eqref{force} and~\eqref{TranslationalDef}.
Here, we set the size ratio} \Ref{to~}\Ref{$\delta = 1$.
The solid lines indicate an estimate of the transition between trapping (blue squares) and penetration (red triangles), given by~$P_0 = 1$.}
}
\label{State-Diagrams}
\end{figure}

\begin{figure}
\begin{center}
 \includegraphics[scale=0.9]{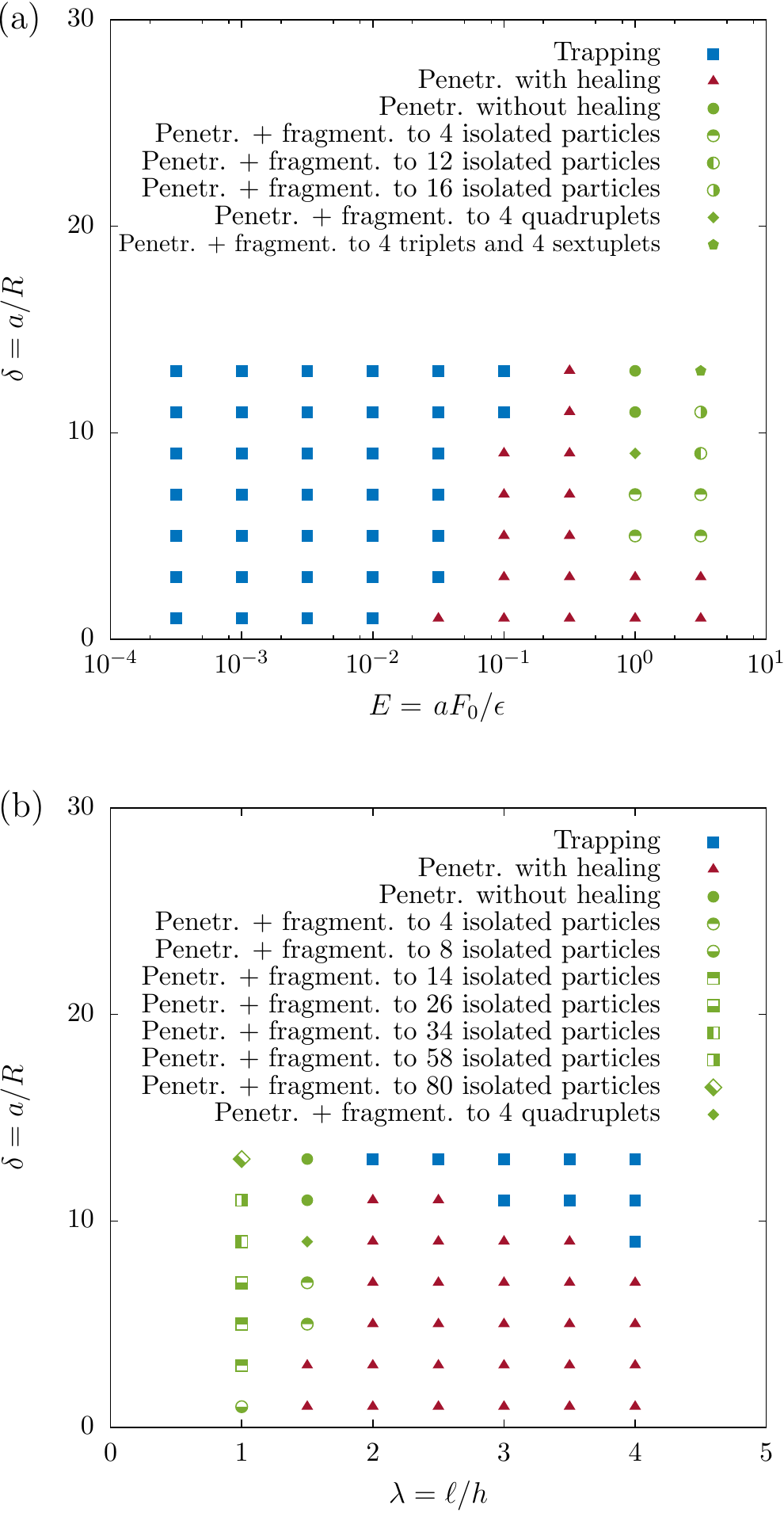}
\end{center}
\caption{\Ref{(Color online) 
State diagram in the parameter spaces~$(a)$~$(\delta, E)$ for $\lambda=3/2$ and~$(b)$~$(\delta,\lambda)$ for $E=1$, where, in both diagrams, $\kappa = 10^{-2}$.
Symbols represent the dynamical state resulting from numerical integration of the governing equations of motion given by Eqs.~\eqref{force} and~\eqref{TranslationalDef}.
In addition to trapping (blue squares) and penetration with healing (red triangles), penetration events without subsequent healing (green symbols) occur in some parameter ranges for large values of the size ratio~$\delta$ and reduced activity~$E$ for~$\lambda \le 3/2$.
These penetration scenarios may be accompanied by the creation of a permanent hole of a size exceeding the interaction range of the membrane particles in addition to the partial fragmentation of the membrane into isolated or clusters of particles.}
}
\label{State-Diagrams-Radius}
\end{figure}

An additional dimensionless parameter, that we denominate as \enquote{admittance}, is introduced to quantify the penetration capability of the active particle.
It is defined based on the above definitions of~$E$ and~$\kappa$ and expresses the ratio between active and elastic forces.
Specifically,
\Ref{
\begin{equation}
	P_0 = \frac{E}{\kappa} = \frac{2F_0}{kh} \label{P0Definition}
\end{equation}
}
Here, the admittance serves to quantify a criterion of whether or not the active passes through the membrane.
For ease of reference, the explicit expressions of the key dimensionless numbers characterizing the states of the system are listed in Tab.~\ref{dimensionlessNumbers}.

To get a first intuition of the possible membrane dynamics, we display the different observed scenarios in Fig.~\ref{IllusStats}.
For low admittance (\Ref{$P_0 = 1$}, top row and movie~S1 in
the Supporting Information), the membrane starts to deform when the
motile active particle comes close, but only up to some point, reaching a
steady state of constant membrane shape and fixed position of the active
particle [see Fig.~\ref{IllusStats}, panels~$(c)$~and~$(d)$].
When increasing the admittance to \Ref{$P_0 = \sqrt{10}$} (second row and movie~S2), the membrane no longer reaches a steady state, but the active
particle breaks through the membrane, leaving a hole that starts to
self-heal once the particle has left the membrane particles behind. 
Here, the
membrane evolves back towards its original configuration, as it would be
desired, e.g., when
delivering cargo to the inside of a healthy cell, the membrane of which we
would want to remain intact. 
The membrane dynamics qualitatively changes when using larger
particles instead \Ref{$(\delta=7)$} and strongly enhancing the admittance to \Ref{$P_0 = 100\sqrt{10}$} (third row and movie~S3).
In this situation, the particle breaks through the membrane and creates a permanent hole.
The four particles located around the center of the membrane in Fig.~\ref{IllusStats}~$(l)$ remain isolated because the range of the internal membrane interactions is shorter than the separation distance of these four particles from the rest of the membrane.
Such a behavior is even more pronounced for significantly larger particles \Ref{$(\delta = 13)$} (bottom row and movie~S4) 
\Ref{where the membrane is partially fragmented into four clusters of particle triplets and four clusters of particle sextuplets (Fig.~\ref{IllusStats}~$(p)$), after the active particle has penetrated through the membrane.}

In Fig.~\ref{State-Diagrams}, we present state diagrams indicating the system behavior \Ref{in the parameter space~$(\kappa, E)$.}
As already mentioned, the membrane is composed of~$N=450$ particles.
Here, we set~$\delta=1$.
Depending on the ratio between \Ref{the control parameters~$\kappa$ and~$E$,} we observe that the active particle either passes through the membrane to reach the other side (red triangles) or remains trapped (blue rectangles).
The transition between the two states can be described by a linear hypothesis \Ref{of the form~$P_0 = 1$.}
Accordingly, penetration events occur when the membrane restoring forces consisting of elastic contributions become weaker than the damaging force resulting from the steric interactions with the active particle.
After full penetration has occurred, the membrane self-heals and relaxes back to its initial equilibrium configuration.

In Fig.~\ref{State-Diagrams-Radius}, we show dynamical state diagrams \Ref{in the planes of the control parameters~$(a)$~$(\delta, E_2)$ for $\lambda=3/2$, and~$(b)$~$(\delta,\lambda)$ for~$E=1$.}
To limit the parameter space, we set in both diagrams the reduced stiffness to~$\kappa = 10^{-2}$.
We observe that the transition between the trapping and penetration states can also be enabled by varying \Ref{the size ratio~$\delta$ (Fig.~\ref{State-Diagrams-Radius}~$(a)$).}
Accordingly, the penetration capability through a membrane is not only determined by the system admittance, but also by the size of the active particle relative to that of the membrane particles.
This is in agreement with earlier experimental investigations indicating that particle size may strongly affect the uptake efficiency and kinetics~\cite{jiang08, andersson11, oh11, huang12, elbakry12}.
Consequently, an active particle with a size larger than that of the membrane particles is more likely to remain trapped.
It is worth noting that, in the considered range of parameters, the transition has been found to  only depend on the admittance~$P_0$ for our simplistic 1D model membrane studied in a previous work~\cite{daddi19penetration}.
For large values of the size ratio \Ref{and small scaled cutoff distance~$\lambda$,} the penetration process may also occur without subsequent self-healing of the membrane.
This situation is accompanied by partial fragmentation of the membrane, during which a number of particles around the center remain isolated \Ref{from the remaining of the membrane,} creating a permanent hole in the membrane.
The number of fragments largely depends on the propulsion speed and the size ratio.
\Ref{Interestingly, the membrane was also observed to become partially fragmented into clusters of four quadruplets for~$(\delta,E) = (9,1)$ or a combination of four triplets and four sextuplets of membrane particles for~$(\delta,E) = (13, \sqrt{10})$.}
This effect points to an interesting size effect of the membrane behavior and shows that motile particles can be used to permanently damage the considered type of membrane.
\Ref{For large scaled cutoff distances~(Fig.~\ref{State-Diagrams-Radius}~$(b)$), the penetration capability decreases, yet the transition between the trapping and penetration states is weakly dependent on the scaled cutoff distance~$\lambda$.
In these situations, the membrane self-heals after particle penetration because the displaced membrane particles remain within the interaction range of the restoring elastic forces.}

\begin{figure}
	\begin{center}
		\includegraphics[scale=0.95]{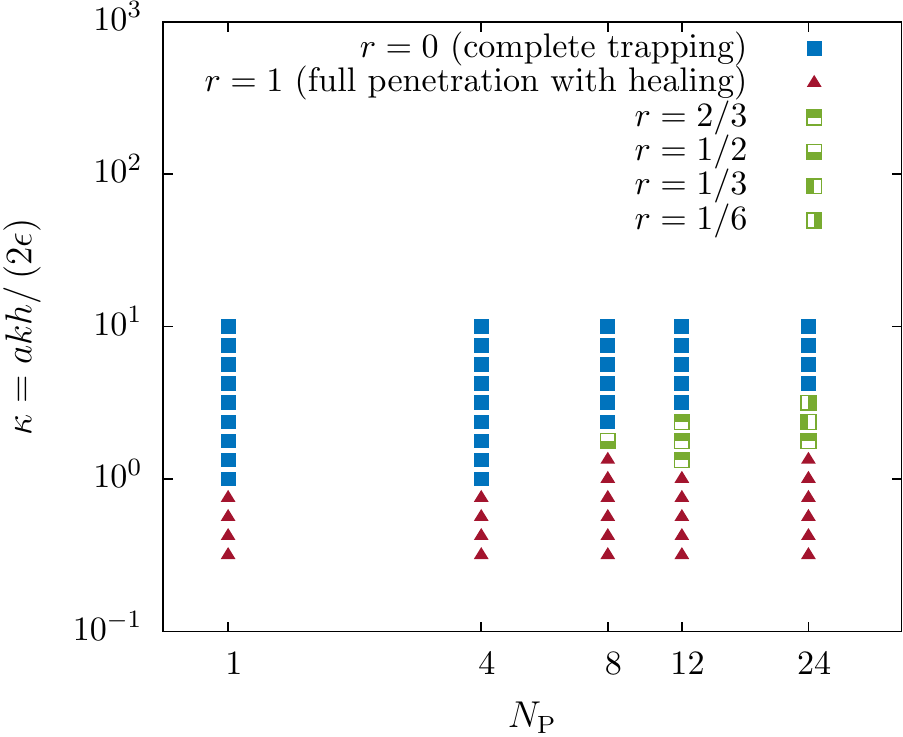}
		\caption{\Ref{(Color online) State diagram in the parameter space~$(\kappa, N_\mathrm{P})$ for~$\delta=1$, $\lambda=3/2$, and~$E=1$.
		Symbols represent the state resulting from numerical integration of the governing equations of motion given by Eqs.~\eqref{force} and~\eqref{TranslationalDef} for~$N_\mathrm{P}$ active particles initially placed around the center of the membrane.
		The parameter~$r$ quantifies the fraction of the active particles that penetrate through the membrane.
		}
		}
		\label{State-Diagrams-NP}
	\end{center}
\end{figure}

\begin{table}
	\def\arraystretch{2.5} 
	\Ref{
	\begin{tabular}{|c|c|}
	\hline
		$N_\mathrm{P}$ & Scaled initial positions~$(x,y)/h$ \\
		\hline
		1 & $(0,0)$ \\
		\hline
		4 & $(\pm \sqt/2, \pm \sqt/2)$ \\
		\hline
		8 & $(\pm \sqt/2, \pm \sqt/2)$, $(0, \pm 1)$, $(\pm 1, 0)$ \\
		\hline
		\multirow{2}{*}{12} & $(\pm \sqt/2, \pm \sqt/2)$, $(0, \pm 1)$, $(\pm 1, 0)$ \\
		   & $(\pm \sqt, \pm \sqt)$ \\
		\hline
		\multirow{3}{*}{24} & $(\pm \sqt/2, \pm \sqt/2)$, $(0, \pm 1)$, $(\pm 1, 0)$ \\
				   & $(\pm \sqt, \pm \sqt)$, $(\pm 2\sqt, 0)$, $(0, \pm 2\sqt)$ \\
				   & $(\pm 3\sqt/2, \pm \sqt/2)$, $(\pm \sqt/2, \pm 3\sqt/2)$ \\
		\hline
	\end{tabular}
	}
	\caption{
	\Ref{Initial Cartesian coordinates of the $N_\mathrm{P}$ active particles in
	the transverse plane}
	}
	\label{table:initPos}
\end{table}

\Ref{
Having investigated the penetration mechanism of a single active particle} \Ref{self-propelling} \Ref{toward the membrane it is worth commenting on the collective penetration of a large number of active particles.
Depending on the density and activity of the penetrating particles as well as on the physical properties of the membrane, the penetration of a group of active particles may show a behavior different from the one observed for a single particle.
In order to probe this effect in some} \Ref{detail, we consider $N_\mathrm{P}$ active particles} \Ref{initially arranged around the center of the membrane.
The active particles move under the action of equal propulsion forces and are} \Ref{initialized} \Ref{at the same vertical distance} \Ref{to} \Ref{the membrane.
The initial Cartesian coordinates of the active particles are listed in Tab.~\ref{table:initPos}.
}

\Ref{
In Fig.~\ref{State-Diagrams-NP}, we present a state diagram in the parameter space~$(\kappa, N_\mathrm{P})$ for~$\delta=1$, $\lambda=3/2$, and $E=1$.
We quantify the proportion of the active particles that pass through the membrane by the ratio~$r$.}
\Ref{Here,} \Ref{$r=0$ corresponds to the situation in which all the active particle are fully trapped while~$r=1$ corresponds to the penetration of all the particles.
The latter scenario is always accompanied by a subsequent self-healing considering the present set of parameters.
We observe that, as the number of active particles gets larger, the penetration capability increases.
This behavior is justified by the fact that, as the number of active particles increases,} \Ref{the forces damaging the membrane} \Ref{become larger than the elastic forces.
Consequently, more active particles are able to break through the membrane \Ref{even if a single particle would be trapped}.
In this context, Kaiser \textit{et al.}~\cite{kaiser12} demonstrated that a chevron-shaped boundary represents an excellent trapping device for self-propelled active particles.
Accordingly, the deformation of the membrane induced by the active particles in the trapping state would eventually trap other particles, thus resulting into an increased penetration capability.}

\section{Analytical theory}
\label{sec:analytischeTheorie}

To rationalize our numerical results, we derive in the following an analytical theory based on a perturbative expansion technique that describes the system behavior in the small-deformation regime.
Particularly, we are interested to determine theoretically \Ref{the membrane displacement field} of the particles in the trapping state.
Our analytical calculations proceed through the linearization of the governing equations of motion, followed by prescribing the relevant fields using a discrete-to-continuum approach~\cite{Goh2018, Menzel2019}.

\subsection{Linearized equations of motion}

In the following, we neglect for simplicity the steric interactions between the membrane particles and assume that the mutual distance between neighboring particle is within the interaction range of the elastic forces, i.e., $r_{ij} \in [h, \ell]$, with $j \in \mathcal{N} (i)$, for~$i=1, \dots, N$.

The dynamical equation governing the evolution of the~$i$th membrane particle displacement field can be cast in the form
\begin{equation} \label{linearizedEqsMotion}
		\dot{\R}_i = \mu \left( \vect{F}_i^\mathrm{E} +  \vect{F}_i^\mathrm{ext} \right)  \, , 
\end{equation}
wherein the superposed dot represents a temporal derivative, and~$\vect{F}_i^\mathrm{E}$ is the elastic force.

Assuming that the active particle has a radius comparable to that of the membrane particles, i.e., for~$\delta\sim 1$, it can readily be verified that the resistive force due to the steric interactions with the active particle vanishes except for the four particles located near the center of the membrane, the initial coordinates of which are given in the Cartesian coordinate system by~$(x,y) = (\pm \sqt h/2, 0)$ and~$(x,y) = (0, \pm \sqt h/2)$.

Following a linear elasticity theory approach~\cite{timoshenko59, sadd09}, we express the position vectors of each particle relative to the laboratory frame as~$\vect{r}_i = \left( U_i + u_i \right) \eX + \left( V_i + v_i\right) \eY + w_i \eZ $, for~$i = 1, \dots, N$, where $U_i \eX + V_i \eY$ is the position vector in the undeformed state of reference, and $u_i \eX + v_i \eY + w_i \eZ$ is the displacement of the membrane particles relative to the initial configuration.
The linearized elastic force acting on the~$i$th particle reads
\begin{widetext}
\Ref{
\begin{equation}
	\vect{F}_i^\mathrm{E} = -\frac{kh}{2}
	\begin{pmatrix}
		2\left(p_1+p_5\right) + \left(2-\xi\right) S_p  + 2\left(1-\xi\right)\left(p_3+p_7\right) + \xi \left( q_2-q_4+q_6-q_8 \right) \\
		\xi \left( p_2-p_4+p_6-p_8 \right)
		+ 2\left(1-\xi\right) \left(q_1+q_5\right)
		+ \left(2-\xi\right)S_q + 2 \left( q_3+q_7 \right) \\
		2\left( 1-\xi \right) \left( S_r + Q_r \right)
	\end{pmatrix} \, .
\end{equation}
}
\end{widetext}
where we have defined~\Ref{$p_j = \left( u_i-u_j \right)/h, q_j = \left( v_i-v_j \right)/h$, and~$r_j = \left( w_i - w_j \right)/h$} to denote the displacement gradients.
Here, the numbers $j = 1, \dots, 8$ appearing in subscript denote the index of a nearest or next-neighbor particle on the lattice, as schematically illustrated in Fig.~\ref{illustration}~$(b)$.
Moreover, we have used the shorthand notations~$S_{\alpha} = \alpha_2+\alpha_4+\alpha_6+\alpha_8$, $Q_{\alpha} = \alpha_1+\alpha_3+\alpha_5+\alpha_7$, for~$\alpha \in \{p,q,r\}$, in addition to $W = p_2+p_4-p_6-p_8+q_2-q_4-q_6+q_8 $.

Notably, the inplane components of the elastic forces involve gradients of the lateral displacements~$p_j$ and~$q_j$.
In contrast to that, the normal components are found to depend on the displacement gradient~$r_j$ only.
Consequently, a decoupling between the lateral and normal displacements is found for planar membranes, in a way analogous to what has previously been observed for 2D elastic membranes that are modeled as a continuum hyperelastic material featuring resistance toward shear and bending~\cite{daddi16b, daddi18stone, daddi18epje, daddi18coupling, daddi19jpsj}.
Particularly, for a non-prestressed membrane~$(\xi=1)$, the elastic forces are purely tangential (oriented along the plane of the membrane) and depend solely on the inplane displacement gradients~$p_j$ and~$q_j$.

Having derived linearized expressions for the forces and torques governing the evolution of the membrane particles, we next consider the dynamics of the active particle.
The latter is subject to the active driving force~$\vect{F}_0 = F_0 \, \eZ$ in addition to the repulsive steric forces resulting from the interaction with the nearby membrane particles.
In the overdamped regime, the translational motion of the active particle along the~$z$ direction is governed by
\begin{equation}
	6 \pi\eta R \, \dot{z}_\mathrm{P} = F_0 - 4\Ref{F^\mathrm{ext}} \cos\alpha \, , 
	\label{GleischungAktivTeilchen}
\end{equation}
wherein~$z_\mathrm{P}$ denotes the $z$-position of the active particle, \Ref{and~$F^\mathrm{ext}$ stands for the magnitude of the steric force exerted by one of the four particles located around the membrane center,} and~$\alpha$ denotes the angle this force makes with the vertical.

Equations.~\eqref{linearizedEqsMotion} form~$2N$ ordinary differential equations in the time variable for the unknown membrane displacement field.
These equations are subject to the initial conditions of vanishing membrane displacement, in addition \Ref{to vanishing displacement} at the membrane periphery and periodic boundary conditions along the~$x$ and~$y$ directions.
In the steady state, the problem is equivalent to searching for the solution of linear recurrence relations \Ref{coupling the positions} of all the membrane particles initially located on a lattice. 
Due to the somewhat complicated nature of the resulting equations, an analytical solution is far from being trivial. 
To handle this difficulty and to obtain a quantitative insight into the system behavior in the small-deformation regime, we will approach the problem differently.
Our solution methodology will be based on a continuum description of the linearized equations of motion as detailed below.

\subsection{Continuum theory}

The core idea of discrete-to-continuum analysis, is to express the membrane displacements following the standard approach as
\Ref{
\begin{equation} \label{disct-to-kontinuum}
	\begin{pmatrix}
		 u_{i+s, i+r} \\
		 v_{i+s, i+r} \\
		 w_{i+s, i+r} 
	\end{pmatrix}
	= \exp \big( h \sqt \left( s D_x + r D_y \right) \big)
	\begin{pmatrix}
		u (x,y) \\
		v (x,y) \\
		w (x,y) 
	\end{pmatrix} ,
\end{equation}
}
where $D_\alpha = \partial / \partial \alpha$, $\alpha \in \{x,y\}$ represents the differential operator and~$(s,r) \in \{ 0, \pm 1/2, \pm 1 \}$.
Here, the fraction at subscripts~$i\pm 1/2$ refer to the nearest-neighboring particle on the lattice axes, namely, the ones identified by even numbers in Fig.~\ref{illustration}~$(b)$.
The integer subscripts~$i\pm 1$ refer to the next-nearest-neighboring particles located on the lattice diagonals.

The exponential argument in Eq.~\eqref{disct-to-kontinuum} can be expanded up to the second order in power series using a two-dimensional Taylor expansion as~\cite{rosenau03}
\begin{equation}
	\begin{split}
			\exp \big( h \sqt ( & s D_x + r D_y ) \big)
					= 1 + h\sqt \left( s D_x + r D_y \right)  \\
					&+ h^2 \left( s^2 D_x^2 + 2sr D_xD_y + r^2 D_y^2 \right) + \cdots  \, .
	\end{split}
\end{equation}

Applying this transformation rule to Eq.~\eqref{linearizedEqsMotion}, the partial differential equation governing the translational degrees of freedom of the membrane particles can be rewritten in vector form as
\begin{widetext}
\Ref{
\begin{equation}\label{translationalDynamik}
	\begin{split}
		\begin{pmatrix}
			u_{,t} \\
			v_{,t} \\
			w_{,t}
		\end{pmatrix}
		 &= \frac{A}{2}
			\begin{pmatrix}
				 	\left(6-\xi\right)u_{,xx} + \left(6-5\xi\right)u_{,yy} + 2\xi \, v_{,xy} \\
				 	\left(6-5\xi\right)v_{,xx} + \left(6-\xi\right) v_{,yy} + 2\xi \, u_{,xy}\\
				 	6 (1-\xi) \left( w_{,xx} + w_{,yy} \right)
				\end{pmatrix} 
				+  h^2 \left( \mu F_0 - \delta \dot{z}_\mathrm{P} \right) \delta (x,y) \, \eZ
			\, ,
	\end{split}
\end{equation}
}\Ref{where~$A:=\mu k h^2$} is a parameter having the dimension of a diffusion coefficient.
Here, we have approximated the steric force exerted on the particles near the center of the membrane by a two-dimensional Dirac delta function~$\delta(x,y)=\delta(x)\delta(y)$.
We have checked that taking alternative forms for the steric force, such as a 2D rectangle function centered around the origin, does not alter our results significantly.
Therefore, a Dirac delta function has been adopted here for simplicity.

In the steady-state limit, Eq.~\eqref{translationalDynamik} simplifies to
\Ref{
\begin{equation}
		\begin{pmatrix}
			\left(6-\xi\right) u_{,xx} + \left(6-7\xi\right)u_{,yy} \\
			\left( 6-7\xi \right) v_{,xx} + \left( 6-\xi \right) v_{,yy} \\
			6 (1-\xi) \left( w_{,xx} + w_{,yy} \right)
		\end{pmatrix} 
	+ 
	h P_0 \, \delta(x,y) \, \eZ = \vect{0}  \, , \label{centralResultOfThePaper}
\end{equation}
}
\end{widetext}
\Ref{wherein $P_0 = 2F_0/(kh)$,} is the system admittance defined above by Eq.~\eqref{P0Definition}.
In Eq.~\eqref{centralResultOfThePaper}, we explicitly observe that~$P_0$ controls how far the active particle can penetrate through the initial planar membrane and leads to a deflection of the membrane.
We note that the 2D Dirac delta function has the dimension of inverse length squared.
In the following, we attempt to obtain closed analytical expressions for the displacement field not only for the steady state but also for transient dynamics situations.

\subsection{Steady solution}

Because of the already-mentioned decoupling between the lateral and normal displacements, the solution for the in- and out-of-plane deformations can be obtained independently.
Since the external force is exerted normal to the plane of the membrane, deformation will predominantly occur along the~$z$ direction.
In the following, we assume that~$|w| \ll L$, for our approximate equations of motion derived above to be valid.

By projecting Eq.~\eqref{centralResultOfThePaper} onto the~$z$ direction, the normal displacement is governed by a second-order partial differential equation of the form
\Ref{
\begin{equation}
	6 \left(1-\xi\right) \left( w_{,xx} + w_{,yy} \right) + h P_0 \, \delta(x,y) = 0 \, , \label{secondOrderPDE_w}
\end{equation}
}

To solve Eq.~\eqref{secondOrderPDE_w}, we exploit periodicity of the system along the transverse directions by expressing the membrane normal displacement~$w$ in terms of a Fourier series~\cite{bracewell99}.
Then,
\begin{equation}
	w(x,y) =  \frac{4}{L^2} \sum_{p \ge 1} \sum_{q \ge 1} \hat{w}(p,q) c_p(x) c_q(y) \, , 
	\label{w_FourierSeries}
\end{equation}
with $p, q = 1, 2, \dots$ denoting the positive integers that set the coordinates in Fourier space.
Here, we have defined the basis function $c_p(x) = \cos \left( H_p x \right)$, where $H_p = (2p-1)\pi/L$, and analogously for~$c_q(y)$.
In addition, $\hat{w} (p,q)$ denotes the Fourier coefficients of~$w$, defined as
\begin{equation}
	\hat{w} (p,q) = \int_{-\frac{L}{2}}^{\frac{L}{2}} \int_{-\frac{L}{2}}^{\frac{L}{2}} 
	                w(x,y) c_p(x) c_q(y) \, \Intd x \, \Intd y \, .
\end{equation}

It is worth mentioning that the solution form given by Eq.~\eqref{w_FourierSeries} follows from the prescribed boundary conditions, so as to ensure that $w(x = \pm L/2, y) = w(x, y = \pm L/2) = 0$.
Moreover, the basis functions~$c_p(x)$ satisfy the orthogonality relation
\begin{equation}
	 \int_{-\frac{L}{2}}^{\frac{L}{2}} c_p(x) c_{p'}(x) \, \Intd x = \frac{L}{2} \, \delta_{pp'} \, .
	 \label{orthogonalityRelation}
\end{equation}

By substituting Eq.~\eqref{w_FourierSeries} into Eq.~\eqref{secondOrderPDE_w} and making use of the orthogonality property given by Eq.~\eqref{orthogonalityRelation}, we readily obtain
\Ref{
\begin{equation}
	\hat{w} (p,q) =  \frac{h P_0}{6 \left(1-\xi\right) \left( H_p^2 + H_q^2 \right)} \, .
	\label{streadyDispl}
\end{equation}
}

Finally, by writing the solution for the transverse displacements~$u(x,y)$ and~$v(x,y)$ in terms of Fourier series in a way analogous to Eq.~\eqref{w_FourierSeries}, it follows that $u$ and $v$ must vanish to satisfy the boundary conditions imposed at the membrane extremities, considering the present approximate equations.

\begin{figure}
\begin{center}
 \includegraphics[scale=0.9]{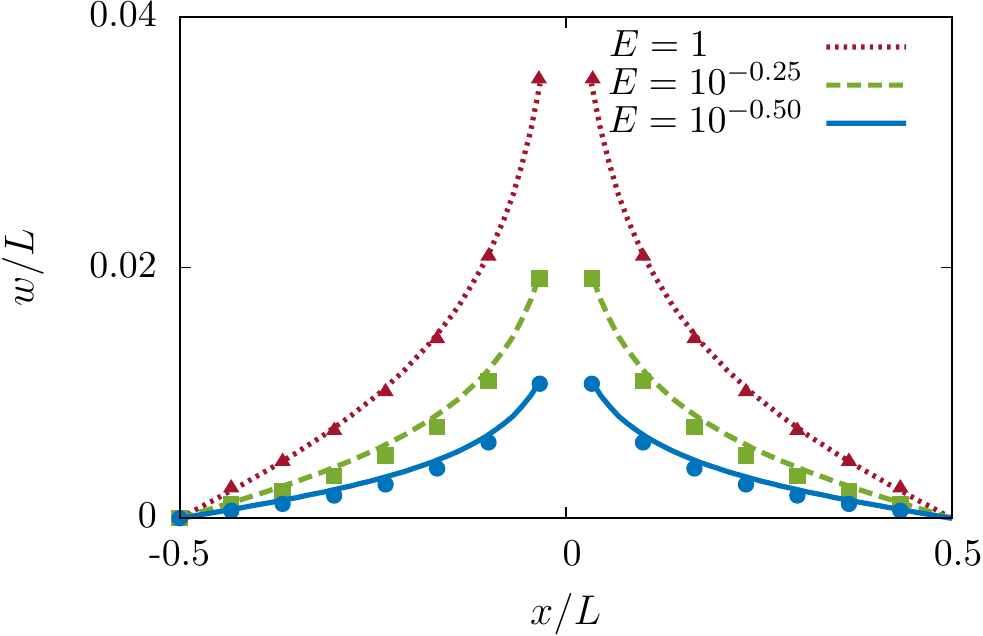}
\end{center}
\caption{(Color online) 
\Ref{Membrane normal displacement} versus scaled distance~$x/L$ calculated in the plane of maximum deformation~$y=0$.
Comparison is shown between analytical predictions (lines) and full numerical solution of Eqs.~\eqref{force} and~\eqref{TranslationalDef} (symbols) in the steady-state of membrane trapping \Ref{for various values of~$E$, while keeping~$\kappa=1$.}
Thus, the elastic interactions have essentially comparable effects on the overall membrane behavior.}
\label{memShape}
\end{figure}

Figure~\ref{memShape} shows the steady-state variations of the normal displacement (scaled by the membrane size) versus~$x/L$.
Results are shown in the plane of maximum deformation~$y=0$ for three different values of the reduced activity~$E$, while keeping the reduced elasticity \Ref{to~$\kappa=1$.}
Symbols indicate the numerical solution of the full nonlinear problem given by Eqs.~\eqref{force} and~\eqref{TranslationalDef} and solid lines are the analytical predictions obtained from the solution of the continuum equations using finite Fourier transforms.
Good agreement is found between the theory and simulations.
All in all, our predictive model requires no fitting parameters and thus can conveniently be applied to describe the steady-state membrane displacement in the small-deformation regime considered here.

\subsection{Transient dynamics}

Having investigated the system behavior in the steady trapping limit, we next turn our attention to the transient dynamics under the action of the force exerted by an active particle pushing against the membrane.
To be able to make an analytical progress, we assume that~$\delta \dot{z}_\mathrm{P} \ll \mu_0 F$, such that~$F_0$ is balanced by the steric interaction with the membrane, not by friction with the fluid.
Accordingly, we set~$\dot{z}_\mathrm{P} = 0$ in Eq.~\eqref{translationalDynamik} for~$t>0$.

The projected equations of motion governing the temporal evolution of the normal displacement field~$w$ read
\Ref{
\begin{equation}
		\frac{w_{,t}}{A} = 3\left(1-\xi\right) \left( w_{,xx} + w_{,yy} \right) + \frac{h P_0}{2} \, \delta(x,y) \, .
\end{equation}
}

Using a similar solution procedure as for the steady dynamics that is based on Fourier transforms, we obtain
\Ref{
\begin{equation}
		\frac{\hat{w}_{,t}}{A} = -3\left(1-\xi\right)\left( H_p^2+H_q^2 \right) \hat{w} + \frac{h P_0}{2} \, .
	 \label{transientFourier}
\end{equation}
}

Applying Laplace transforms~\cite{widder15} to Eqs.~\eqref{transientFourier} and solving for the unknown field~$\hat{w}$, we readily obtain
\Ref{
\begin{equation}
		\hat{w}(p,q) = \frac{h P_0 A}{2s \left( 3A\left(1-\xi\right) \left(H_p^2+H_q^2\right) + s \right)} \, .
\end{equation}
}

\Ref{
The expressions of the Fourier coefficient in the time domain follow forthwith by inverse Laplace transform as
\begin{equation}
	\frac{\hat{w}}{\hat{w}_\infty}  =  1 - e^{-3A\left(1-\xi\right) \left(H_p^2+H_q^2\right) t} \, , \notag
\end{equation}
}
where~$\hat{w}_\infty$ represents the steady normal displacement given by Eq.~\eqref{streadyDispl}.

\begin{figure}
\begin{center}
 \includegraphics[scale=0.9]{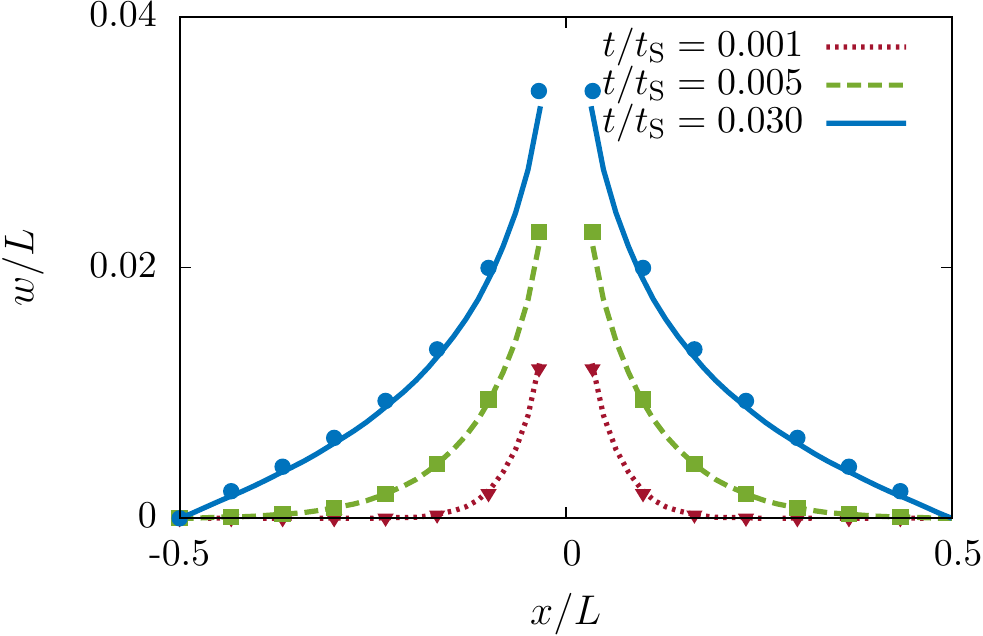}
\end{center}
\caption{(Color online) 
\Ref{Membrane normal displacement} out of the plane of the undeformed membrane as functions of scaled distance~$x/L$ calculated in the plane~$y=0$.
Comparison is made between analytical predictions (lines) and full numerical simulations (symbols) for the transient behavior before the trapping state at various scaled times where, again,~$\tS = \eta L^3/\epsilon$ denotes the unit of simulation time.
\Ref{Here, we set $(\kappa, E) = (1,1)$.}}
\label{memShapeTransit}
\end{figure}

In Fig.~\ref{memShapeTransit}, we present the transient evolution of the membrane normal displacement before reaching the steady-state at three scaled times, where~$t_\mathrm{S} = \eta L^3/\epsilon$ denotes the simulation time.
Here, curves are shown in the plane~$y=0$ \Ref{using the membrane parameters~$(\kappa, E) = (1,1)$.
As time increases, the membrane deformation exponentially approaches the steady-state value.}
Although the analytical theory involves no fitting parameters, very good agreement is obtained between full numerical simulations (symbols) and analytical predictions (lines).

\section{Conclusion}
\label{sec:conclusions}

In the present work, we have discussed the interaction of an active
particle with a minimal 2D membrane which could be realized, e.g.,
using synthetic particles of controlled interactions.
We have identified three different scenarios, one corresponding to a
permanent trapping of the particle by the membrane and the remaining two
implying penetration of the particle through the membrane. 
The
first type of penetration is characterized by a complete subsequent healing of the
membrane which relaxes towards its equilibrium configuration once the
particle has passed. In stark contrast, we have shown that much larger
particles can create a hole in the membrane that is large enough to
prevent such a self-healing dynamics, resulting in a permanently
damaged membrane. 
This behavior is accompanied by the expulsion of membrane particles into isolated fragments.
Our result suggests that if one were to
effectively damage a synthetic vesicle, or perhaps a cancer cell
membrane, one would need to use particles of a certain minimal size.
Complementary to simulations, we here
provide a detailed analytical theory allowing to predict the entire
state diagram, the shape and the dynamics of the membrane.
Our approach might be useful to predict transitions between trapping,
penetration with and without self-healing in experiments.

\begin{acknowledgments}
	We thank Joachim Clement and Cornelia Monzel for stimulating discussions.
	The authors gratefully acknowledge support from the DFG (Deutsche Forschungsgemeinschaft) through the projects DA~2107/1-1, ME~3571/2-2, and LO~418/16-3.
\end{acknowledgments}

%

\end{document}